\documentclass[aps,prl,floats,twocolumn,showpacs,superscriptaddress]{revtex4}
\usepackage{graphics}
\usepackage[pdftex]{graphicx}
\usepackage{amsmath,bm}
\usepackage{relsize}
\begin{document}

\title{Synchronization of oscillators through time-shifted common inputs}

\author{Ehsan Bolhasani}
\affiliation{Department of physics, Institute for Advanced Studies in Basic Sciences, Zanjan, Iran}
\affiliation{School of Cognitive Science, Institute for Research in Fundamental Sciences (IPM), P. O .Box 1954851167, Tehran, Iran}

\author{Yousef Azizi}
\affiliation{Department of physics, Institute for Advanced Studies in Basic Sciences, Zanjan, Iran}

\author{Alireza Valizadeh}
\affiliation{Department of physics, Institute for Advanced Studies in Basic Sciences, Zanjan, Iran}
\affiliation{School of Cognitive Science, Institute for Research in Fundamental Sciences (IPM), P. O .Box 1954851167, Tehran, Iran}

\author{Matja{\v z} Perc}
\affiliation{Faculty of Natural Sciences and Mathematics, University of Maribor, Koro{\v s}ka cesta 160, SI-2000 Maribor, Slovenia}
\affiliation{CAMTP -- Center for Applied Mathematics and Theoretical Physics, University of Maribor, Krekova 2, SI-2000 Maribor, Slovenia}

\begin{abstract}
Shared upstream dynamical processes are frequently the source of common inputs in various physical and biological systems. However, due to finite signal transmission speeds and differences in the distance to the source, time shifts between otherwise common inputs are unavoidable. Since common inputs can be a source of correlation between the elements of multi-unit dynamical systems, regardless of whether these elements are directly connected with one another or not, it is of importance to understand their impact on synchronization. As a canonical model that is representative for a variety of different dynamical systems, we study limit-cycle oscillators that are driven by stochastic time-shifted common inputs. We show that if the oscillators are coupled, time shifts in stochastic common inputs do not simply shift the distribution of the phase differences, but rather the distribution actually changes as a result. The best synchronization is therefore achieved at a precise intermediate value of the time shift, which is due to a resonance-like effect with the most probable phase difference that is determined by the deterministic dynamics.
\end{abstract}


\maketitle
\section{Introduction}
Synchronization has been observed in a wide variety of physical and biological systems, and it is indeed a universal phenomenon in nonlinear sciences \cite{pikovsky2003synchronization, strogatz2003sync}. The functionality of systems often depends on the degree of synchronization, with examples including power grid systems~\cite{blaabjerg2006overview, rohden2012self, motter2013spontaneous}, Josephson junction arrays~\cite{barbara1999stimulated, likharev1986dynamics}, cardiac pacemaker cells~\cite{jalife1984mutual, peskin1975mathematical}, and gamma rhythms in nervous systems~\cite{gray1989stimulus, rabinovich2006heteroclinic, besserve2015shifts}, to name but a few examples. The subject is an evergreen field of research in statistical physics, attracting recurrent attention in relation to chaotic systems, neuronal dynamics, and network science \cite{pecora1990synchronization, rulkov1995generalized, rosenblum1996phase, neiman1999noise, boccaletti2002synchronization, arenas2006synchronization, arenas2008synchronization, fischer2006zero, gomez2007paths, gomez2011explosive, aguirre2014synchronization}.

The coordination of activity of nonlinear oscillators is conventionally assigned to the presence of attractive interactions between them~\cite{kuramoto2003chemical}. However, synchronization among coupled oscillators in the presence of delayed interactions is more challenging to understand~\cite{pyragas1998synchronization, yeung1999time, lee2009large, wang2009synchronization, sadeghi2014synchronization}. Delayed interactions between coupled oscillators are in general attributed to finite information transmission speeds. As has been shown before, such interactions can change attractive coupling to a repulsive coupling and vice versa, and thus crucially affect the emergence of synchronization in networks of coupled oscillators~\cite{dhamala2004enhancement, pototsky2009delay, sen2010effect, d2014stochastic, esfahani2016stimulus}.

But regardless of presence of direct interactions between the oscillators, synchronization is still attainable through a common input. The presence of such inputs evokes statistical correlations between the oscillators, which can ultimately lead to synchronization~\cite{goldobin2005synchronization, teramae2004robustness, pikovsky1992statistics, yu1990transition, bolhasani2014direct, hauschildt2006noise,kawamura2016optimization}. In studies of the synchronization through common inputs, signal transmission delays can be ignored if the common source affects the target oscillators through equidistant routes, but of course not if the distances to the downstream oscillators differ. In the latter case, the input cross-correlation function for every pair of oscillators peaks at a different time that depends on the specific time shift. As yet, this realistic setup constitutes an important unsolved problem in the realm of synchronization among nonlinear oscillators.

In this paper, we therefore study the emergence of phase-synchronization among limit-cycle oscillators that are driven by time-shifted common inputs. In particular, we determine the effects on the output cross-correlation function of two oscillators when they receive time-shifted common inputs and the time shift is smaller than the period of oscillation of the oscillators. If the oscillators are uncoupled, we show that the effect of a time-shifted common input is simply a shift in the output cross-correlation function. For coupled oscillators, however, time shifts in the common inputs non-trivially affect the output correlation so that in addition to the shift, the correlation function and the distribution function for the phase difference of the oscillators are dependent on the input time-shift. We show analytically and confirm numerically that the degree of synchrony depends on the time-shift of common inputs, such that optimal synchronization is obtained only when the common inputs are differentially shifted by a non-zero time lag.

\section{Mathematical framework}

\begin{figure*}
\centering
\includegraphics[width=0.8\textwidth ]{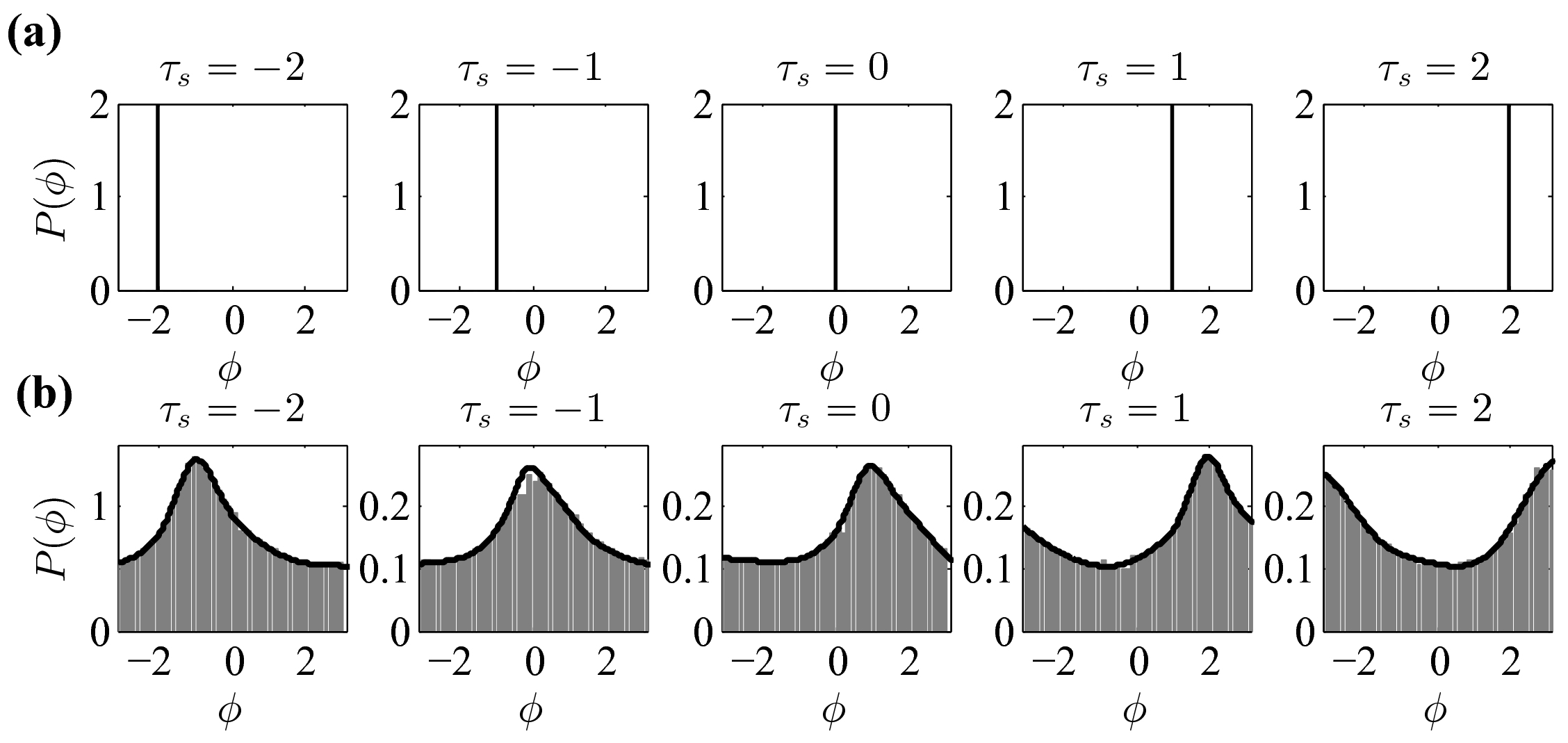}
\caption{(a) Phase difference distribution function of two identical uncoupled oscillators, receiving a fully correlated Gaussian white noise with different time shifts shown above the plots. (b) Same results for non-identical oscillators with the frequency mismatch $\Delta\omega = 0.5$. Here $D=0.5$, and $\tau _{s}$ is $-2, -1, 0, 1, 2$ from left to right.}
\label{fig1}
\end{figure*}

We consider two bidirectionally coupled oscillators with state vector $X_i$, $(i=1,2)$ which evolves according to
\begin{eqnarray}
&\dot{{\bm X}}_{1}(t)& = {\bm F}\Big({\bm X}_{1}(t),I_{1}\Big) \nonumber\\ &+&\hskip-0.38cm \varepsilon g_{12}{\bm G}_{12}\Big({\bm X}_{1}(t), {\bm X}_{2}(t)\Big) + \sqrt{D\varepsilon} {\bm \xi} (t),\nonumber\\
&\dot{{\bm X}}_{2}(t)& = {\bm F}\Big({\bm X}_{2}(t),I_{2}\Big) \nonumber\\ &+&\hskip-0.38cm \varepsilon g_{21}{\bm G}_{21}\Big({\bm X}_{2}(t), {\bm X}_{1}(t)\Big) + \sqrt{D\varepsilon} {\bm \xi} (t-\tau _{s}),\nonumber\\
\end{eqnarray}
where vector function $F(X_{i}, I_{i})$ describes the inherent dynamic of the isolated oscillators ($g_{ij}=0$). We assume that the isolated oscillators have a stable limit cycle $X_{LC}(t) = X_{LC}(t+T)$ with the period $T$ which is controlled by the scalar parameter $I_i $. $G_{ij}$ determines the interaction function whose strength is defined by $g_{12}$. The last terms in the equations describe common stochastic input to the two oscillators where $\xi (t)$ is Gaussian white noise with zero mean and unit variance and $D$ determines the noise intensity. $\tau _{s}$ is the key parameter of the present study which is assumed to be smaller than the period of the oscillation of the two oscillators ($\tau_s < T$) and determines the time shift between the two stochastic inputs, i.e., the second neuron receives exactly the same stochastic input to the first neuron, but with a time shift equal to $\tau _{s}$. Both the interaction term and stochastic inputs are scaled with the small parameter $\varepsilon \ll 1$. Furthermore the oscillators are assumed to be in general nonidentical due to the small difference in their control parameter $I_1 -I_2 =\varepsilon\Delta I$.

Applying the standard phase reduction method in the regime of weak perturbation \cite{kuramoto2003chemical, teramae2004robustness} leads to the following $\rm{It{\hat{o}}}$ stochastic differential equations, for the time evolution of the phase variable $\theta(X)$ in the vicinity of the unperturbed limit cycle $X_{LC}$
\begin{eqnarray}
\dot{\theta}_1 (t) = \omega_1 &+&\varepsilon  g_{12}{\bm Z}\left( \theta _{1}\right) \cdot {\bm G}\left(\theta _{1}(t), \theta _{2}(t) \right)\nonumber\\ &+& \sqrt{D \varepsilon } {\bm Z}\left( \theta _{1}\right) \cdot {\bm \xi} \left( t \right)  \nonumber \\
\dot{\theta}_2 (t) = \omega_2 &+&\varepsilon  g_{21}{\bm Z}\left( \theta _{2}\right) \cdot {\bm G}\left(\theta _{2}(t), \theta _{1}(t) \right)\nonumber\\ &+& \sqrt{D \varepsilon } {\bm Z}\left( \theta _{2}\right) \cdot {\bm \xi} \left( t -\tau _{s}\right)
\label{ophase}
\end{eqnarray}
where $Z(\theta)=\nabla_X \theta|_{X_{LC(\theta)}} $ is the phase sensitivity \cite{winfree2001geometry}. Since we assumed small inhomogeneity in the bifurcation parameter $\Delta I \sim \mathcal{O} (\varepsilon )$, difference in natural frequencies will be also small $\omega _{1} - \omega _{2} = \varepsilon\Delta\omega$.

\begin{figure*}
\centering
\includegraphics[width=0.8\textwidth ]{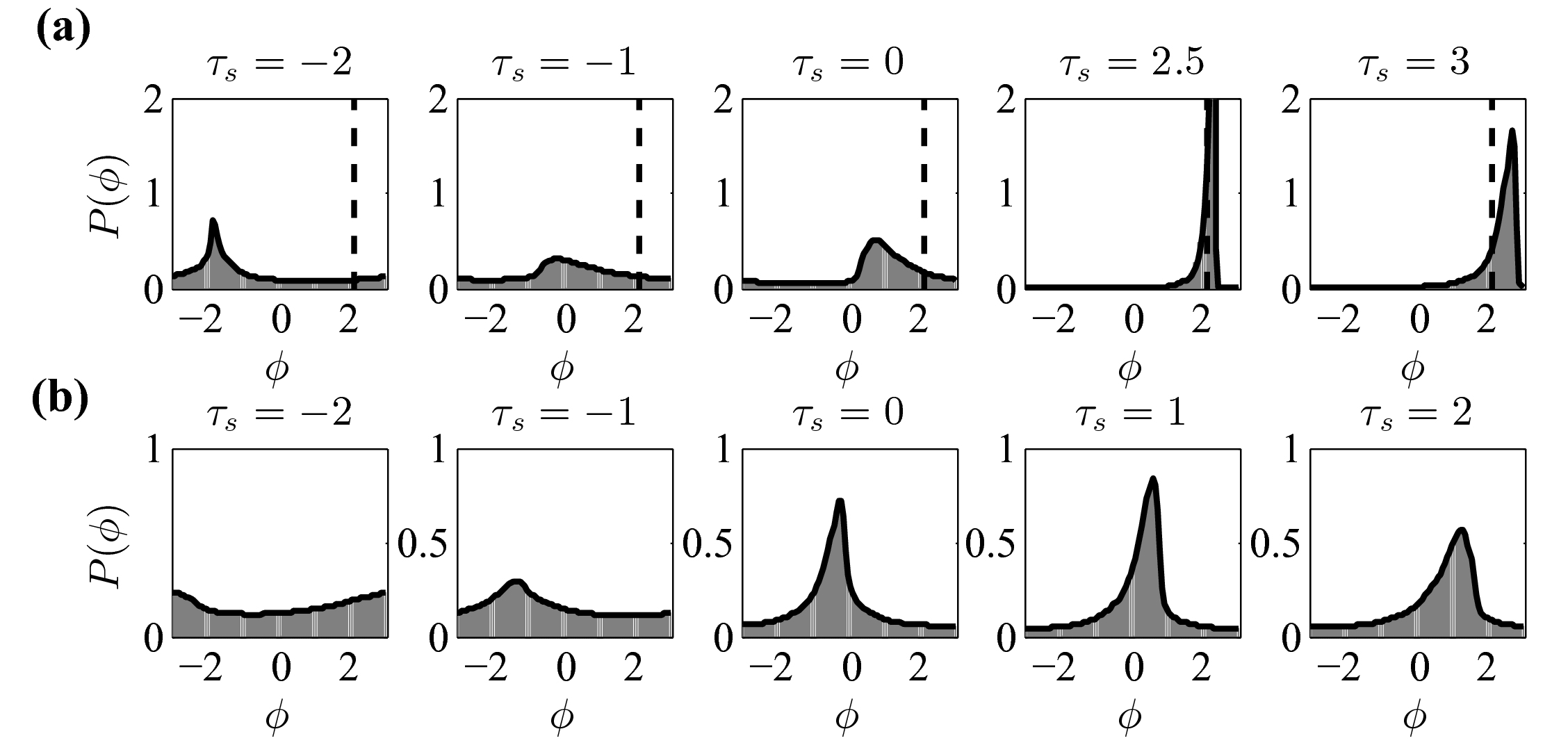}
\caption{(a) Phase difference distribution function of two coupled oscillators in the locked mode, receiving a fully correlated Gaussian white noise with different time shifts shown above each plot. (b) Same results for oscillators in the running mode. The results shown by gray bar plots are calculated by numeric integration of Eq.~3 and the solid lines show the analytical result given by Eq.~6. The vertical dashed lines in (a) show the fixed point of Eq.~3. Other parameters are $D=0.5$ and $\Delta\omega=0.2$ for the locked case, and $\Delta\omega=-0.1$ for the running case.}
\label{fig2}
\end{figure*}

We take $\omega_1=1$ and $\omega_2=1-\varepsilon\Delta \omega$ and define time-shifted phase difference as $\theta (t)=\theta _{1}(t) - \theta _{2}(t +\tau _s)$. From Eqs. \ref{ophase} we obtain the evolution of $\theta (t)$ to the order of $\varepsilon$
\begin{eqnarray}
\dot{\theta}(t)&&= \varepsilon\Delta\omega +\varepsilon \Big( H_{12}(\theta _{1}(t), \theta _{2}(t)) \nonumber\\ &&-  H_{21}(\theta _{2}(t+\tau _{s}), \theta _{1}(t+\tau _{s})) \Big) \nonumber\\
&&+\sqrt{\varepsilon}{\bm f}\Big(\theta _{1}(t), \theta _{2}(t+\tau _{s})\Big)\cdot {\bm \xi}(t)\nonumber\\
\end{eqnarray}
here
\resizebox{0.40\textwidth}{!}{${\bm f}\Big(\theta _{1}(t), \theta _{2}(t')\Big)=\sqrt{D} \left[ {\bm Z}\Big( \theta _{1}(t)\Big)-{\bm Z}\Big( \theta _{2}(t')\Big)  \right]$} and
$H_{ij}(\theta _{i}(t), \theta _{j}(t))=g_{ij}{\bm Z}\Big( \theta _{i}(t)\Big)\cdot {\bm G}\left(\theta _{i}(t), \theta _{j}(t) \right)$. We assume $\theta_i (t) = t + \varphi_i(t)$ where the first term $t$ captures the intrinsic dynamics of isolated oscillators and the second term is slow varying deviation from natural oscillations. Averaging the equation over one period \cite{ermentrout1991multiple}, we have
\begin{eqnarray}
	\hspace{-30pt}\dfrac{d\varphi (t)}{dt} = \varepsilon\Delta\omega &+&\varepsilon \Big( \bar{H}_{12}(\varphi)-\bar{H}_{21}(-\varphi) \Big)
	\nonumber\\
	&+&\sqrt{\varepsilon}\bar{f}\Big(\varphi-\tau _{s}\Big)\eta(t)
	\label{phi_2}
\end{eqnarray}
where $\varphi=\varphi _{1} -\varphi _{2}$ and the averaged functions $\bar{H}_{ij}$ and $\bar{f}$ are defined as $\bar{H}_{ij}(\theta _{i}(t), \theta _{j}(t))=\frac{1}{2\pi}\int _{0}^{2\pi} dt H_{ij}(\theta _{i}(t), \theta _{j}(t))$ and  $\bar{f}(\varphi- \tau _s)=\sqrt{\dfrac{1}{2\pi}\int _0^{2\pi} dt \Big[f(\theta _1(t), \theta _2(t+\tau _s ))\Big]^2}$ (see Appendix). Note that in the above equation the lag in the common inputs $\tau_s$ acts like delay in the connections with a change of variable $\varphi \rightarrow \varphi+\tau_s$. We finally derive the Fokker-Planck equation for the distribution of the phase differences of the two oscillators, described by Eq.~\ref{phi_2}
\begin{eqnarray}
\frac{\partial\rho}{\partial t} \left( \varphi , t \right) = &-&\varepsilon\frac{\partial}{\partial \varphi} \left[ \Gamma \left( \varphi \right) \rho \left( \varphi , t \right) \right] \nonumber \\ & + &  \varepsilon  \frac{\partial ^{2}}{\partial \varphi ^{2}} \Big( \bar{f}(\varphi - \tau _{s})  \rho\left( \phi , t \right)\Big)
\label{Fokker}
\end{eqnarray}
where $\Gamma \left( \varphi \right)=\Delta \omega+\Big( \bar{H}_{12}(\varphi)-\bar{H}_{21}(-\varphi)\Big)$ and $\rho (\varphi,t)$ depicts the distribution of $\varphi$. The stationary distribution of the phase differences can be calculated by letting $\partial\rho / \partial t=0$
\begin{eqnarray}
\resizebox{0.40\textwidth}{!}{$\rho_0(\varphi)=\dfrac{e^{M(\varphi )}}{N\bar{f}(\varphi - \tau _{s})}\left[ \dfrac{e^{-M(2\pi)}-1}{\int _{0}^{2\pi}{e^{-M(x)}dx}}\int _{0}^{\varphi}e^{-M(x)} dx +1\right]$}
\label{stationary}
\end{eqnarray}
where $M(\varphi) = \int _{0}^{\varphi} \frac{\Gamma (x)}{\bar{f}(x - \tau _{s})}dx$ and $N$ is a normalization factor.
In the following the analytical results are obtained by calculation of the stationary distribution function from the above equation (see Appendix for the details of derivations).

\section{Results}
To check the validity of the Eq.~4 and the corresponding solution of Fokker-Planck equation Eq.~6, we take $Z(\theta ) = 1-\cos (\theta )$ which is the canonical form of the phase sensitivity for the type-I oscillators near SNIC bifurcation \cite{izhikevich2007dynamical}. Furthermore, we assume the oscillators are pulse-coupled, i.e., $G_{ij} = \Sigma _{n}\delta\left( t-t_{j}^{n}\right)$ in which $\delta $ is Dirac's delta function and $t_{j}^{n}$ is the instant of $\theta_j=2 \pi n$. Pulse coupling approximation for interaction between oscillators is valid in the systems where the interaction term activates over a time which is small compared to period of the oscillation \cite{gerstner1996rapid, hansel2001existence, peskin1975mathematical, strogatz2001exploring}. To assess the degree of synchrony regardless of the value of the phase lag we use the synchronization index $\gamma ^2 = \langle\cos \varphi \rangle ^2 + \langle\sin \varphi \rangle ^2$, where the brackets denote the averaging over time \cite{tass1998detection}.

\begin{figure}
\centering
\includegraphics[width=0.47\textwidth ]{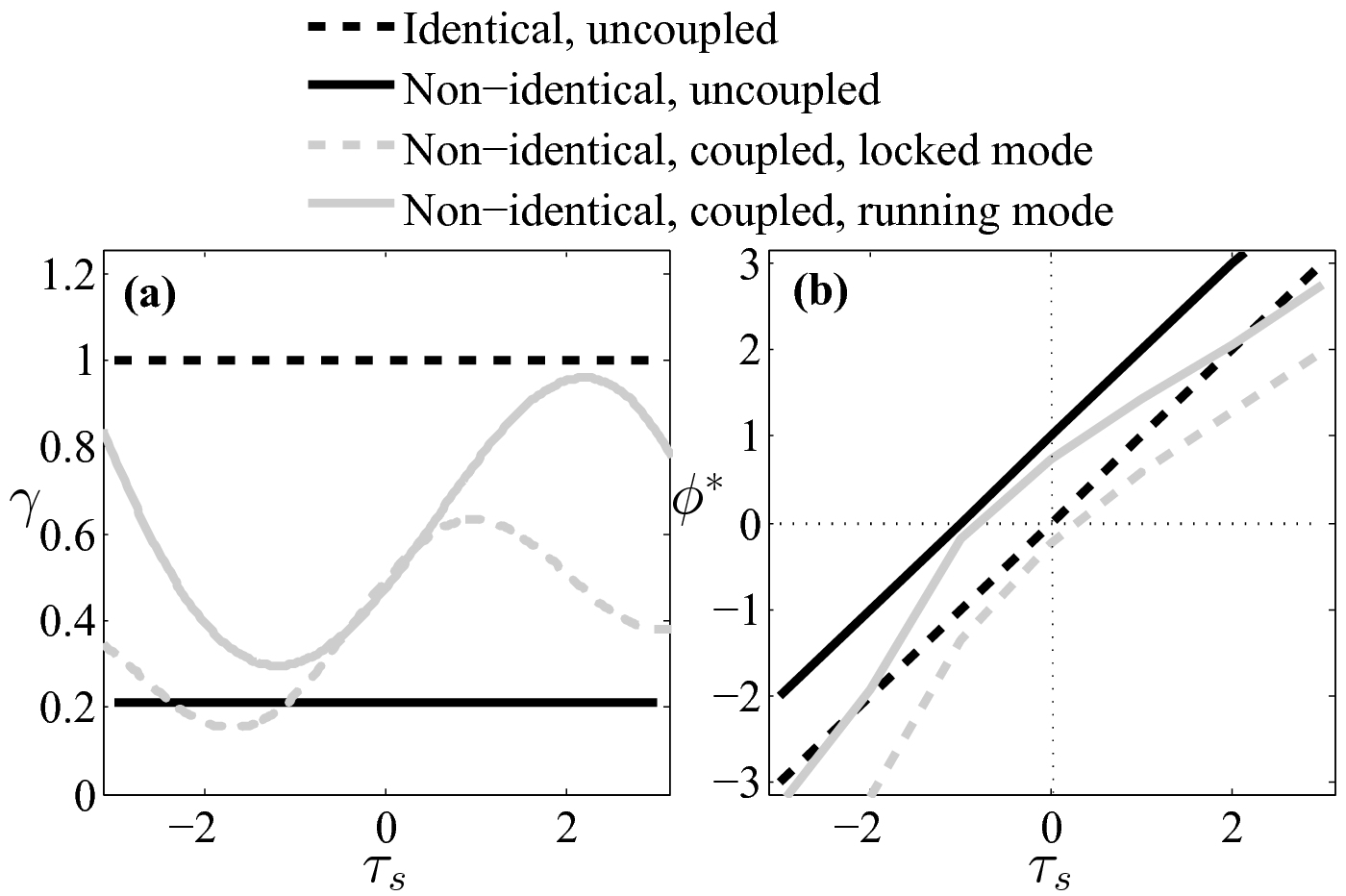}
\caption{(a) Synchronization index versus input time-shift for the case of two identical uncoupled phase oscillators (black dashed line), non-identical uncoupled phase oscillators (solid black line), coupled phase oscillators in phase-locked mode (solid gray line), and non-identical coupled phase oscillators in the running mode (gray dashed line). (b) Most probable phase lag which shows the location of the peak of the phase lag distributions in Figs.~1 and 2 is plotted versus input time shift.}
\label{index}
\end{figure}

First, we considered two uncoupled oscillators receiving a common noise with a time shift $\tau_s$. For the identical oscillators $\Delta \omega =0$, it has been shown that the oscillators synchronize when receiving common noises \cite{teramae2004robustness}. Our results show that time shift in the inputs results in the same time shift in the synchronized output without changing the synchrony index (Fig.~\ref{fig1}a and Fig.~\ref{index}). When the oscillators are not identical $\Delta \omega \ne 0$, phase difference distribution function spreads and synchrony index decreases (Fig.~\ref{fig1}b and Fig.~\ref{index}). In the presence of heterogeneity (mismatch of the firing rates) the most probable phase lag $\phi^*$ is not zero when the inputs have zero time-shift \cite{burton2012intrinsic}. Interestingly this effect of heterogeneity can be compensated by a non-zero time shift in the inputs, i.e., the most probable phase difference of the oscillators could be around zero despite the heterogeneity by a suitable choice of the time lag of inputs (see Fig.~1b with $\tau_s \simeq -1$ and Fig.~\ref{index}b). This means that the maximum zero-lag correlation of the two non-identical oscillators is achieved when the input to the high-frequency oscillator is lagged. Note that in this case changing the lag in the inputs does not change the functional form of the distribution function and so the synchrony index.

Unlike the case of uncoupled oscillators the effect of time lag in the inputs to the coupled oscillators is not restricted to the shift of the distribution of the phase lags. For a system of two coupled oscillators, two cases can be recognized. In the first case (locked mode) the deterministic version of Eq.~4 (with no stochastic input) has a stable fixed point, and in the second case there is no fixed point for Eq.~4 and the system is in the {\it running mode}.
The effect of common inputs in these two cases are shown in Fig.~\ref{fig2}a and b, respectively. In the locked mode the synchrony index is no longer independent of the time lag of the inputs and peaks when the time lag matches the phase lag of deterministic case (Fig.~\ref{index}). Accordingly, while the location of the peak of the distribution function shifts with changing input time lag, its maximum value and the width of the distribution around the most probable phase lag also changes (Fig.~\ref{fig2}a). Analytical results obtained by the solving the stationary Fokker-Planck equation (Eq.~6) confirm the results of the simulation.

In the running mode the results are qualitatively similar to the locked mode except for the overall decrease in the synchrony index and wider distribution of the phase differences (Fig.~\ref{fig2}b and Fig.~\ref{index}). Changing the time-lag of the inputs the distribution is shifted while its width and maximum are changed. Again, the best entrainment with the maximum synchrony index is attained in a certain value of time lag of the inputs ($\tau_s \simeq 1$, see Fig.~\ref{fig2}b and Fig.~\ref{index}). Analytical results from Eq.~6 are again validated by the direct numeric solution of Eq.~3.

\section{Summary}
Summarizing, we have studied the emergence of synchronization between two limit-cycle oscillators subject to a common but time shifted stochastic input. In addition to showing that for uncoupled oscillators the effect of time shifts is a trivial corresponding shift in the distribution of the phase differences between the two oscillators, we have derived fundamental conditions for optimal synchronization when the oscillators are coupled. Namely, we have shown analytically that the time shift between the inputs changes the distribution of relative phases, and with it also the degree of synchronization in the system. Specifically, with time shifts that are in accord with the most probable phase difference between the oscillators due to their deterministic dynamics, a resonance-like effect can be observed that leads to optimally phase locked oscillators. Due to the generality of the considered setup, we expect our results to significantly improve our understanding of phase synchronization in networks of nonlinear elements, especially within the realm of multilayer networks, where common inputs across different layers might be particularly likely \cite{boccaletti2014structure, kivela2014multilayer}.

\appendix*
\setcounter{equation}{0}
\renewcommand{\theequation}{A\arabic{equation}}
\section{Appendix}

We define $\theta (t) = \theta _{1}(t) - \theta _{2}(t+\tau _{s})$ and subtract the equations Eq.~2 after shifting the time in second equation by $\tau_s$ to obtain the evolution equation for $\theta (t)$ (Eq.~3). Then we define
\begin{eqnarray}
\theta _{i}(t) = t+\varphi _{i}(t)
\end{eqnarray}
where the first term in R.H.S. captures the intrinsic dynamics of isolated oscillators with $\omega_i \simeq 1$, and the second term is slow varying deviation from natural oscillations.
Using the method of averaging \cite{ermentrout1991multiple}, we substitute Eq.~A1 into Eq.~3 and average Eq.~3 over one period. For the second term in the R.H.S. of Eq.~3 we define
\begin{eqnarray}
\bar{H}( \varphi )&=&\dfrac{1}{2\pi}\int _0^{2\pi}H( \theta _{1}(t), \theta _{2}(t))dt\nonumber\\
&=&\dfrac{1}{2\pi}\int _0^{2\pi}H( t+\varphi _{1}, t+\varphi _{2} )dt\nonumber\\
&=&\dfrac{1}{2\pi}\int _0^{2\pi}H( \bar{t}, \bar{t}-\varphi )d\bar{t}
\end{eqnarray}
and for the third term in R.H.S. of Eq.~3 we should take into account the integral of the correlation term as follows
\begin{widetext}
\begin{eqnarray}
\dfrac{1}{2\pi}\int _0^{2\pi}\int dt~dt' \Big\langle\Big[{\bm f}(\theta _1(t), \theta _2(t)+\tau _s )\cdot{\bm \xi}(t)\Big] \times \Big[{\bm f}(\theta _1(t'), \theta _2(t')+\tau _s )\cdot{\bm \xi}(t')\Big]\Big\rangle \nonumber
\end{eqnarray}
\begin{eqnarray}
=\sum _{ij}\dfrac{1}{2\pi}\int _0^{2\pi}\int dt~dt' \Big\langle\Big[{\bm f}_{i}(\theta _1(t), \theta _2(t)+\tau _s ){\bm \xi} _{i}(t)\Big]\Big[{\bm f}_{j}(\theta _1(t'), \theta _2(t')+\tau _s ){\bm \xi}_{j}(t')\Big]\Big\rangle \nonumber
\end{eqnarray}
\begin{eqnarray}
=\sum _{ij}\dfrac{1}{2\pi}\int _0^{2\pi}\int dt~dt' \Big[{\bm f} _{i}(\theta _1(t), \theta _2(t)+\tau _s )\Big]  \Big[{\bm f}_{j}(\theta _1(t'), \theta _2(t')+\tau _s )\Big]\Big\langle{\bm \xi}_{i}(t){\bm \xi}_{j}(t')\Big\rangle\nonumber
\end{eqnarray}

\begin{eqnarray}
=\sum _{i}\dfrac{1}{2\pi}\int _0^{2\pi}\int dt~dt' \Big[{\bm f}_{i}(\theta _1(t), \theta _2(t)+\tau _s )\Big] \Big[{\bm f}_{i}(\theta _1(t'), \theta _2(t')+\tau _s )\Big]\delta(t-t')\nonumber
\end{eqnarray}
\end{widetext}
\begin{eqnarray}
=\dfrac{1}{2\pi}\int _0^{2\pi} dt \Big[{\bm f}(\theta _1(t), \theta _2(t)+\tau _s )\Big]^2
\label{fs}
\end{eqnarray}
\begin{eqnarray}
=\dfrac{1}{2\pi}\int _0^{2\pi} dt \Big[{\bm f}(t+\varphi _1, t+\tau _s+\varphi _2 )\Big]^2\nonumber
\end{eqnarray}
\begin{eqnarray}
=\dfrac{1}{2\pi}\int _0^{2\pi} d\bar{t} \Big[{\bm f}(\bar{t}, \bar{t}-\varphi  +\tau _s)\Big]^2.
\label{ff}
\end{eqnarray}

According to Eqs.~\ref{fs} and \ref{ff} we can define
\begin{eqnarray}
\Big[\bar{f}(\varphi- \tau _s)\Big]^2=\dfrac{1}{2\pi}\int _0^{2\pi} dt \Big[f(\theta _1(t), \theta _2(t)+\tau _s )\Big]^2\nonumber\\
\label{S10}
\end{eqnarray}
and using Eq.~3 and Eq.~A2  to \ref{S10} we obtain the evolution equation for $\phi=\phi_1-\phi_2$
\begin{eqnarray}
\dfrac{d\varphi (t)}{dt} = \varepsilon\Delta\omega &+&\varepsilon \Big( \bar{H}(\varphi)-  \bar{H}(-\varphi) \Big)\nonumber\\&+&\sqrt{\varepsilon}\bar{f}\Big(\varphi-\tau _{s}\Big)\eta(t)
\label{phi_3}
\end{eqnarray}
where $\eta (t)$ is another Gaussian white noise with zero mean and unit variance.
In a more compact form we would have
\begin{eqnarray}
\dfrac{d\varphi (t)}{dt} = \varepsilon\Gamma(\varphi )
+\sqrt{\varepsilon}\bar{f}\Big(\varphi-\tau _{s}\Big)\eta(t)
\label{phi_4}
\end{eqnarray}
where
\begin{eqnarray}
\Gamma (\varphi ) = \Delta\omega + \Big( \bar{H}(\varphi)-  \bar{H}(-\varphi) \Big).
\end{eqnarray}

The Fokker-Planck equation for the phase difference distribution of Eq.~\ref{phi_4} is
\begin{eqnarray}
	\frac{\partial\rho}{\partial t} \left( \varphi , t \right) = &-&\varepsilon\frac{\partial}{\partial \varphi} \left[ \Gamma \left( \varphi \right) \rho \left( \varphi , t \right) \right] \nonumber\\&+& \varepsilon  \frac{\partial ^{2}}{\partial \varphi ^{2}} \Big( \bar{f}(\varphi - \tau _{s})  \rho\left( \phi , t \right)\Big)
	\label{Fokker}
\end{eqnarray}
and the steady state solution of this equation can be achieved by solving $\partial \rho (\varphi , t) / \partial t = 0$. Defining
\begin{eqnarray}
R(\varphi )=\bar{f}(\varphi - \tau _{s})  \rho\left( \phi , t \right)
\end{eqnarray}
we can rewrite Eq.~\ref{Fokker} for $\bar{f}(\varphi - \tau _{s})\neq 0$ as
\begin{eqnarray}
	\frac{\partial R}{\partial t} \left( \varphi , t \right) = &-&\varepsilon\frac{\partial}{\partial \varphi} \left[ \dfrac{\Gamma \left( \varphi \right)}{\bar{f}(\varphi - \tau _{s})} R \left( \varphi , t \right) \right]\nonumber\\ &+& \varepsilon  \frac{\partial ^{2}}{\partial \varphi ^{2}}   R\left( \phi , t \right).
\label{FokkerR}
\end{eqnarray}
General solution of Eq.~\ref{FokkerR} is
\begin{eqnarray}
R(\varphi ) = \dfrac{1}{N}e^{M(\varphi )}\Big[ (A\int ^{\varphi} e^{-M(x)} dx)+1 \Big]
\end{eqnarray}
where $N$ and $A$ are constant which can be derived by normalization and periodicity conditions, and
\begin{eqnarray}
 M(\varphi )= \mathop{\mathlarger{\int}} ^{~~~~~\varphi } dx \dfrac{\Gamma \left( x \right)}{\bar{f}(x - \tau _{s})},
\end{eqnarray}
and then we derive steady state phase difference distribution $\rho_0(\varphi)$ as Eq.~6 .

To check the validity of the results we take
\begin{eqnarray}
Z(\theta ) = 1-\cos (\theta )
\end{eqnarray}
which is the canonical form of the phase sensitivity for the type-I oscillators near SNIC bifurcation \cite{izhikevich2007dynamical} and assumed that oscillators are pulse-coupled
\begin{eqnarray}
G_{ij} = \Sigma _{n}\delta\left( t-t_{j}^{n}\right)
\end{eqnarray}
substituting these equations in equations A1 to A8 results
\begin{eqnarray}
\Gamma (\varphi) =\dfrac{g_{12}-g_{21}}{2\pi}\Big( 1-cos(\varphi)\Big)
\label{phi_3}
\end{eqnarray}
and
\begin{eqnarray}
\bar{f}(\varphi - \tau _{s})=\sqrt{\Big(1-cos(\varphi -\tau _{s})\Big)}.
\end{eqnarray}
We have then solved the integrals of Eq.~6 numerically to obtain the analytical result for steady state phase difference distribution which is presented in Fig.~2 of the manuscript.

\section{Acknowledgments}

This research was supported by the Iranian National Elites Foundation, by the Cognitive Sciences and Technologies Council, and by the Slovenian Research Agency (Grant Nos. J1-7009 and P5-0027).


\end{document}